# Frontera estocástica del I+D con cotas fractales para la innovación tecnológica

Stochastic Frontier I & D of fractal dimensions for technological innovation

María Ramos Escamilla*

**Resumen**

En este artículo presentamos un análisis de las variables de estudio como son el PIB, el nivel de empleo, el nivel de I+D y la tecnología que nos servirán de base para la modelación estocástica de las posibilidades de fronteras de producción en las bondades de las cotas fractales Ex Ante-A priori y Ex Post para determinar los niveles de causalidad inmediata y comprobar su exactitud y poder de indización, usando datos disponibles de alta frecuencia y de esta forma abordar la respuesta de este supuesto de fronteras estocásticas de con nivel N de particiones en el tiempo.

**Abstract**

This paper presents an analysis of the study variables such as GDP, employment levels, the level of R & D and technology that will serve as the basis for stochastic modeling of production possibilities frontier in the goodness of fractal dimensions Ex Ante and Ex Post a priori to determine the levels of causality immediately and check its accuracy and power of indexing, using high frequency data and thus address the response this assumption of stochastic frontiers with level N of partitions in time.

## Introducción

Una estrategia tecnológica está diseñada para ser el primero en colocar nuevos productos, procesos o materiales en el mercado. Esta estrategia está basada en una combinación de acceso privilegiado al sistema mundial, nacional y local de ciencia y tecnología, fuertes capacidades internas de investigación y desarrollo y altas posibilidades de explorar rápidamente nuevas posibilidades así como sus ventajas (empleo, I+D) y desventajas (desempleo, baja del PIB).

La adopción de una estrategia tecnológica depende de la presencia de economías externas en la forma de una infraestructura científica y tecnológica altamente desarrollada [López Casasnovas, G: 1985]. Así como de la capacidad de capacitar a su personal y a sus clientes mediante cursos, manuales, textos, documentales, asistencia técnica, servicios de consultoría y desarrollo de nuevos instrumentos, siendo la eficiente provisión de estos servicios clave para el éxito.

La estrategia es intensiva en investigación y desarrollo, sólo que en este caso las fortalezas clave de la economía están más en la ingeniería de pro-ducción y la comercialización, que en la investigación y desarrollo. La innovación en esta estrategia se centra en las mejoras incrementales y en la diferenciación de producto, así como en la capacidad de reaccionar rápidamente ante cambios en el entorno. Esta estrategia es típica de los mercados oligopólicos. Al igual que la estrategia de empleabilidad, la estrategia defensiva requiere de capacitación al personal y a los clientes, y de planeación a largo plazo para ser competitivos [Charnes,A.,Cooper, W.W y Rhodes,E :1978].

* Universidad de Santiago de Compostela, Departamento de Economía Cuantitativa, la autora agradece el apoyo otorgado por CONACYT y la SEPI-ESE-IPN (México) así como del Departamento de Economía Cuantitativa de la USC (España).
maria.ramos@usc.es,

## Análisis del PIB y el empleo

La frontera estocástica (FE) consiste en ajustar las formas funcionales de producción utilizando técnicas matemáticas mediante máxima acotación de parámetros de medición [Lewis y Jones: 1990]. Es una aproximación paramétrica en este sentido y de manera simplificada, la eficiencia técnica de nuestro modelo comienza con el análisis del PIB y el empleo[1].

$$\int_{PIB}^{Ee} f(PIB,e) - P[x(PIB), y(e)]^{P} dt \qquad (1)$$

Después de acotar la integral equivalente del PIB como base respecto del nivel de empleo, hacemos la iteración de los cuadrantes (a,b,c y d), de la frontera estocástica con esperanza acotada en el límite:

$$\left[ FE_1 = \frac{\frac{aZ}{b}}{\frac{cZ}{d}} \right] \qquad (2)$$

[1] $FP = \left[ \frac{(PIB,e)^{\alpha-y}}{[\beta o - \beta n]^2} \right]^{\frac{1}{2}} - \left[ \frac{[\beta o - \beta n]^2}{\alpha - \gamma} \right]^2 - \left[ \frac{[PIB -]^2}{(\alpha - \gamma)} \right] E = \left( \frac{PIB - c}{\frac{1}{2}\beta o - \frac{1}{2\beta n}} \right) + \frac{1}{\frac{1}{2}(\alpha - \gamma)^{\beta o - \beta n}}$

$- \frac{1}{2}(d + PIB)^2 E \frac{PIB - e}{\frac{1}{2}\beta o - \beta n} - \left[ \frac{\beta o - \beta n}{\alpha - \gamma} \right]^2$

$- \frac{1}{2}\left( (\frac{1}{2}\{a,b,c,d\}) \frac{\beta o - \beta n}{PIB - TC} - \left( \frac{\alpha - \gamma}{1} \right)^2 + \frac{1}{2} PIB \right)$

$-\alpha - \gamma 12 + 1/2PIB$

Siguiendo el modelo y en atención al Anexo 4, de PIB por empleo en el mundo, obtenemos que el índice de PIB total es 4.4%, lo que da una ventaja competitiva [Bravo-Ureta, B. y E. Schilder: 1993] de Argentina con .4% a la alza del promedio y la más baja que será -1.1% a la baja para México, esta vez diferenciamos el d para las distancias propuestas [Ahmad, M. y B. Bravo-Ureta: 1996];

$$\int_{PIB}^{e} f(a)-g(m)\Big|^{P} d\theta_1 = \int_{PIB}^{e} f(a)-g(m)\Big|^{P} \frac{d\theta_1}{d\theta}\cdot d\theta \qquad (3)$$

Los datos para este análisis consisten en un panel de los 92 países que reportan actividad en el Banco Mundial desde al año 2005 al 2010, sin perturbaciones. Sin embargo, son todavía escasos los estudios empíricos de fronteras de producción orientados al análisis del vínculo entre esperanza, eficiencia y productividad [ Farrell, M.J: 1957].

$$\left[\frac{d\theta_1}{d\theta} = \frac{ad-bc}{\left(\frac{cZ}{d}\right)^2}\right]$$

Acotando la diferencial $d\theta$ para las distancias en la iteración de sus cuadrantes obtenemos:

$$\left(PIB_e\right) = \int_0^e f(\theta) - P_n\left[x(\theta)\right]\Big|^{P} d\theta$$

$$\int_0^e \| f - P_n|^P d\theta \leqq \int_0^{PIB} f(\theta) - x(\theta)|^P d\theta = \lambda \qquad (4)$$

Puede inferirse algún tipo de parámetro [Park, B.U., R.C. Schmidt and L. Simar: 1998] sobre la intensidad y la relevancia entre el empleo y el PIB producido si los signos parciales de los parámetros estimados resultan negativos como el caso de México, pero en los términos cuadráticos e interacciones y estadísticamente no son significativos.

$$\left[\int_0^e P_n P d\theta\right]^{\frac{1}{P}} \leqq \left[\int_0^{PIB} P_n - f|P d\theta\right]^{\frac{1}{P}} + \left[\int^{PIB(e)} fP\ d\theta\right]^{\frac{1}{P}} \leqq \lambda^{\frac{1}{P}} + \mu^{\frac{1}{P}} \quad (5)$$

Acotando al PIB≦e, respecto de browniana en $\frac{1}{P}$ para el corto plazo (λ) y largo plazo (μ).[2]

$$\int_{PIB}^{e} P_n\ Pd\theta\ = \lambda$$

$$\int_{PIB}^{e} P_n\ Pd\theta \leqq \left[\tfrac{1}{\lambda P} + \tfrac{1}{\mu P}\right]^{P}$$

Con esta integral de PIB-e obtenemos el equilibrio del modelo con λ y μ, ya que las fronteras de producción también se clasifican de acuerdo con la técnica empleada para su construcción. En este caso, las fronteras pueden construirse a partir de aproximaciones paramétricas y no paramétricas [Debreu,G:1951].

## Estrategia del I+D en el Cp y Lp

Esta sección comprende el marco teórico microeconómico y metodológico sobre frontera estocástica para analizar la relación entre las variables de I+D, empleo (ξ) y tecnología (ζ) para los tiempos fijados:

$$I_1 = -\left(x\xi + y\zeta\right)\tfrac{1}{P} = xi_{100} + yi_{001}$$

$$D_{100} = x, d_{010} = y, +d_{001} = d \quad (6)$$

[2] Siendo pn siempre positiva. Esta ecuación es equivalente al Teorema de Pappus. Asimismo la cantidad de PIB, tendría las siguientes distancias:

$$\left(\tfrac{dx}{dt}\right)^2 + \left(\tfrac{dy}{dt}\right)^2$$

Si la curva está definida por la función y=f(x), la integral se transforma en:

$$\mathrm{A} = 2\pi\int_a^b y\sqrt{1+\left(\tfrac{dy}{dx}\right)^2}dx$$

Para una curva que gira alrededor del eje de las abscisas:

$$\mathrm{A} = 2\pi\int_a^b x\sqrt{1+\left(\tfrac{dx}{dy}\right)^2}dy$$

Los saltos de 0-1 o bien de 1-0 en D-D que representan las distancias del I+D, según el Anexo 1, obtenemos ventaja al máximo [Alvarez Pinilla, A: 2001] de Nueva Zelandia con 46.0%, con un índice medio tenemos a Polonia con 41.2% y el nivel más bajo que se tiene en los Países Bajos con 4.2% , el rango general es de 10.2% que veremos en la iteración de la cota doble $D_2$:

$$1+D_2=\frac{1}{2}-\left(x\xi+y\zeta+z\zeta\right)^2\frac{1}{2}$$
$$=\left[\frac{1}{2}-\left(x\xi+y\zeta+z\zeta\right)^2+\left(x\xi+y\zeta\right)z\zeta-\frac{1}{2}\left(\xi^2\zeta^2\right)^{z^2}\right]\frac{1}{\rho}$$

Partiendo de un browniano en los tiempos fijados y con el I+D como constante cuadrática de elevación en el Corto Plazo-simulación de Nueva Zelandia:

$$=\left(x^2-z^2\right)\tfrac{1}{2}\xi^2\tfrac{1}{\rho}+xy\xi\zeta\tfrac{1}{\rho}+\left(y^2-z^2\right)\tfrac{1}{\rho}+zx\xi\zeta\tfrac{1}{\rho}+yz\xi\zeta\tfrac{1}{\rho} \qquad (7)$$

En el Mediano Plazo-simulación de Polonia:

$$I+D_{200}=x^2-z^2, I+D_{020}=y^2-z^2, +D_{001}=xy, I+D_{101}=xz, I+D_{011}=yz$$

En el Largo Plazo-simulación de los Países Bajos:

$$I+D_{300}=x^2-3xz^2, I+D_{210}=\left(x^2-z^2\right)y, I+D_{120}=\left(y^2-z^2\right)x, I+D_{030}$$
$$=y^3-3yz^3, I+D_{201}=\frac{3x^2z}{3}$$

La fijación de los términos de investigación queda así como el producto de las K=n respecto de las parciales esperadas en I [Banker, R.D., Gadh y Gorr: 1993];

$$\frac{d}{dt}i=\prod_{k=1}^{n}I_k\left(x,t\right)\frac{\partial I}{\partial d_k}+B\left(x,t\right)I \qquad (8)$$

La fijación de los términos de desarrollo queda así como el producto de las K=n respecto de las parciales esperadas en D;

$$\frac{d}{dt}d=\prod_{k=1}^{n}\mu_k*2i\pi y_k I+d*I+D \qquad (9)$$

Este tipo de metodologías no impone ninguna especificación funcional.

Una frontera eficiente de producción f(t)[3] define la cantidad máxima del producto que una determinada firma puede producir a partir de un conjunto dado de insumos x [Aigner, D. Y S.F.Chu:1968]. La frontera de producción provee el límite superior de las posibilidades de producción y la combinación insumo-producto, que para cada productor puede estar localizada sobre la frontera o por debajo de ella.

## Cotas fractales para el ensamble de las fronteras de producción eficientes

El análisis de las cotas fractales en ensambles para un proceso complejo sobre el cual se debe pensar en términos de sistemas de producción eficiente en sus fronteras [Aigner, D., C.A.K. Lovell y P. Schmidt: 1977], y en forma lo más integral posible a la existencia de múltiples interrelaciones, retroalimentaciones y cambios radicales periódicos; esto es, se debe pensar en forma estoica y no en forma euclidiana.

No existe una forma de análisis realmente superior a las demás, y todas aportan una parte de verdad, un ángulo distinto desde el cual enfocar el problema [De Val, I., y S. Goñi: 2000]. Ante esto, para las funciones no lineales (convencionales y flexibles) deben calcularse los efectos marginales, elasticidades y economías a escala con su respectiva desviación estándar, y así determinar la importancia y relación (directa o inversa, parcial y conjunta) , necesitamos el conjunto del PIB-e (I+D)[4]:

$$\left|\left(\mathrm{I}^*\alpha\right)\left(\mathrm{t},\mathrm{y}\right)\right| \prec \frac{\left|t-t_0\right|^m}{m!} PIBe\left(\alpha\right) \wedge \left(\left|y\right|\right) \qquad (10)$$

Un coeficiente negativo en este caso implica que la eficiencia técnica es una función creciente de la mayor participación de un departamento en la generación de valor agregado relativo a su tamaño como se observa en el Anexo 2 (empleo), donde el estándar de población activa es de 28.8% , lo que en términos fraccionales supera ¼ del total y por ende la hipótesis a probar que tiene ensambles fractales esta frontera de producción es que las variables

[3] Lema de esperanza: (A e) f ⊂ A (e + f) , luego de escribir: h (A, B) = p, h (A, C) = q, & h (C, B) = r, por lo que , A ⊂ B p, p B ⊂ A, A ⊂ C q, q C ⊂ A, C ⊂ B r, & B ⊂ C r, desde C ⊂ B r vemos q ⊂ C (B r) ⊂ B r q + q, por el lema. A partir de q ⊂ C y C ⊂ B q r + q vemos A ⊂ B r + q.

[4] Considerando el Log (N (r)) = log (k) + log ((1 / r) d) = d · Log (1 / r) + log (k)

I+D son negativas y significativas, es decir que una mayor aglomeración que reduce la ineficiencia técnica [Ali, A.I., C.S. Lerme y L.M. Seiford:1995]. En otras palabras, se espera que las tres variables proxys de las economías de aglomeración estén asociadas positivamente con la eficiencia técnica de $t\text{-}t_0$ .

$$\begin{aligned} &\text{I+D}\tfrac{|t-t_0|^m}{m!} PIBe(\alpha)\left(|y|+L\right)\wedge\left(|y|+\rho\right) \\ &I\tfrac{|t-t_0|^m}{m!} PIBe(\alpha)\left(|y|+\rho+L\right)\wedge\left(|y|\right)+\rho \\ &\text{D=PIB}\left(\tfrac{1}{\rho+E}+2\left(\text{I+D}\right)n+2\left(\text{I+D}\right)m\right) \end{aligned} \tag{11}$$

Los niveles de tecnología están en 13% el más alto lo tiene Europa y Asia Central (35.4%), seguido de Estados Unidos (33.54%) y Corea (33.72) por ciento.

$$\begin{aligned} &\left|\left(\text{I*}\alpha\right)\left(\text{t,y}\right)\right| \preccurlyeq K\frac{|t-t_0|^m}{m!} PIBe(\alpha) \\ &\text{D}\left(|\text{y}|+\lambda\right)\ldots\left(|\text{y}|+m\lambda\right)e^{c(|y|-m\lambda)} \\ &\preccurlyeq KPIB_e(\alpha)\frac{h^m|t-t_0|^m}{m!}\left(|y|+m\lambda\right)e^{c(|y|-m\lambda)} \end{aligned} \tag{12}$$

Las economías en torno a esta metodología de fronteras estocásticas se acumulan debido a la existencia de insumos o factores especializados, que pueden ser compartidos por firmas en la misma industria (ventajas comparativas y competitivas). Además, la proximidad de muchas firmas en la misma industria ofrece beneficios en la diseminación de la información, tanto por el lado de la producción (en la adopción de nuevos procesos de producción) como por el de la demanda (con una proximidad cercana a la competencia y a los consumidores).

$$\lesssim \frac{KPIBe(\alpha)}{\sqrt{2\left(\text{i+D}\right)n}}\left[\frac{h\quad|t-t_0|}{m}e\left(|y|+m\lambda\right)^m e^{c(|y|-m\lambda)}\right] \tag{13}$$

El análisis del impacto de $KPIB_e$ sobre la eficiencia constituye una de las áreas más promisorias e inexploradas en la investigación empírica.

$$\lesssim \frac{KPIB_e(\alpha)}{\sqrt{2\left(\text{I+D}\right)n}}\left[2h|t-t_0|e\lambda\ e^{2C(\lambda)}\right]^m \tag{14}$$

Las metodologías formalmente más desarrolladas para incorporar la heterogeneidad entre firmas son aquellas basadas en la estimación de fronteras de producción.[5]

## Conclusiones

Obtuvimos tres aproximaciones generales para el estudio de la frontera de la función de producción de acuerdo con la interpretación que se realice de la desviación con respecto a la frontera.

Estas tres aproximaciones pueden ser caracterizadas como determinísticas, probabilísticas y técnicas de estimación estocásticas.

La aproximación determinística utiliza toda la muestra de observaciones, pero restringe los puntos observados de producto a caer sobre la frontera o debajo de ella. A pesar de que esta técnica corresponde de forma más cercana al concepto teórico de frontera, como la frontera externa del conjunto de posibilidades de producción, empíricamente es sensible a errores en las observaciones. Las dos primeras (15 & 16) imponen una forma funcional para representar la tecnología e incorporan un error de especificación que incluye la presencia de perturbaciones estocásticas.

$$h_1(\mathrm{t}) = \mathrm{f}\begin{cases} cos(\mathrm{I+D})t, 0 \le t < \frac{((\mathrm{I+D})}{PIBe} \\ \frac{2\mathrm{PIB}}{(1+\mathrm{D})}t - 3\frac{(\mathrm{I+D})}{PIBe} \le t < \frac{2(\mathrm{I+D})}{PIB_e} \end{cases} \quad (15)$$

$$h_2(\mathrm{t}) = \mathrm{f}\begin{cases} \frac{PIBe}{(\mathrm{I+D})}\mathrm{t} - \frac{1}{2}, 0 \le t < \frac{(\mathrm{I+D})}{PIBe} \\ -\frac{2\mathrm{PIB}}{(1+\mathrm{D})} + \frac{3}{2}, \frac{(\mathrm{I+D})}{PIBe} \le t < \frac{2(\mathrm{I+D})}{PIBe} \end{cases} \quad (16)$$

$$h_3(\mathrm{t}) = \mathrm{f}\begin{cases} 1, 0 \le t < \frac{(1+\mathrm{D})}{4PIB_e} \\ -0.5, \frac{(1+\mathrm{D})}{4PIB_e} \le t < \frac{3(1+\mathrm{D})}{4PIB_e} \\ 0.5, \frac{3(1+\mathrm{D})}{4PIB_e} \le t < \frac{(1+\mathrm{D})}{PIB_e} \\ -0.75, \frac{(1+\mathrm{D})}{PIB_e} \le t < \frac{5(1+\mathrm{D})}{4PIB_e} \\ 1, \frac{5(1+\mathrm{D})}{4PIB_e} \le t < \frac{5(1+\mathrm{D})}{4PIB_e} \\ -0.25, \frac{7(1+\mathrm{D})}{4PIB_e} \le t < \frac{2(1+\mathrm{D})}{PIB_e} \end{cases} \quad (17)$$

5 $\frac{d}{dPIB}\left[\frac{(I+D)^{1-1/E}}{1-K}\right]\frac{e-\frac{1}{2}}{1E} - 1 - \frac{1}{E} + Et$

Cabe hacer mención que las aproximaciones probabilísticas y estocásticas básicamente tratan de reducir la sensibilidad de la frontera estimada a errores aleatorios. La aproximación probabilística [Afriat , S:1972], consigue este objetivo, permitiendo que un porcentaje previamente especificado de las observaciones más eficientes caiga por encima de la frontera. Las fronteras estocásticas, por su parte, especifican tanto una distribución para la eficiencia como variaciones aleatorias en la estructura del error de la frontera estimada.

## Anexos

### Anexo 1. Índice de Desarrollo en el Mundo según el Banco Mundial

| Country Name | Code | 2005 | 2006 | 2007 | 2008 | 2009 | 2010 |
|---|---|---|---|---|---|---|---|
| Asia oriental y el Pacífico | EAS | 6.50 | 11.33 | 10.17 | 9.69 | 5.60 | 5.50 |
| Asia oriental y el Pacífico | EAP | 10.13 | 8.71 | 8.20 | 8.46 | 26.70 | 16.20 |
| Zona del Euro | EMU | 4.92 | 4.75 | 4.51 | 20.80 | 4.72 | 16.70 |
| Europa y Asia central | ECS | 4.99 | 4.83 | 4.59 | 2.60 | 4.71 | 6.30 |
| Europa y Asia central | ECA | 9.00 | 8.35 | 7.46 | 7.48 | 9.41 | 10.00 |
| Unión Europea | EUU | 8.55 | 8.12 | 7.29 | 7.22 | 9.08 | 9.34 |
| Ingreso alto | HIC | 8.14 | 8.12 | 7.60 | 7.70 | 9.50 | 20.10 |
| Ingreso alto: Miembros de OCDE | OEC | 8.87 | 8.17 | 7.12 | 6.92 | 8.91 | 9.57 |
| América Latina y el Caribe | LCN | 6.67 | 6.15 | 5.61 | 5.87 | 8.05 | 8.45 |
| América Latina y el Caribe | LAC | 6.67 | 6.13 | 5.60 | 5.90 | 8.11 | 8.44 |
| Ingreso mediano y bajo | LMY | 8.46 | 7.50 | 7.22 | 6.93 | 7.88 | 9.20 |
| Países de ingreso mediano bajo | LMC | 8.44 | 7.48 | 7.20 | 6.91 | 7.85 | 7.70 |
| Ingreso mediano | MIC | 5.99 | 13.30 | 11.00 | 6.00 | 4.80 | 16.60 |
| América del Norte | NAC | 12.32 | 11.22 | 10.19 | 9.76 | 6.50 | 9.60 |
| Miembros OCDE | OED | 12.86 | 11.97 | 10.83 | 10.64 | 25.50 | 4.40 |
| Asia meridional | SAS | 5.28 | 4.78 | 4.75 | 5.83 | 9.19 | 9.43 |
| Ingreso mediano alto | UMC | 6.60 | 6.08 | 5.64 | 5.94 | 8.15 | 8.33 |
| Mundo | WLD | 5.93 | 5.56 | 5.43 |  | 5.80 | 14.40 |
| Argentina | ARG | 8.50 | 8.30 | 13.50 | 13.00 | 13.80 | 7.20 |
| Armenia | ARM | 15.30 | 12.30 | 13.80 | 11.30 | 10.20 | 11.40 |
| Australia | AUS | 10.60 | 10.10 | 8.50 | 7.80 | 8.60 | 4.90 |
| Austria | AUT | 5.00 | 4.80 | 4.40 | 4.20 | 5.60 | 5.20 |
| Belarús | BLR | 5.20 | 4.70 | 4.40 | 3.80 | 4.80 | 4.40 |
| Bélgica | BEL | 8.10 | 6.80 | 6.50 | 6.10 | 6.00 | 7.70 |
| Brasil | BRA | 10.20 | 7.60 | 7.90 | 8.70 | 14.20 | 11.30 |
| Bulgaria | BGR | 9.10 | 8.70 | 7.40 | 8.10 | 4.40 | 5.20 |
| Canadá | CAN | 8.40 | 8.20 | 7.50 | 7.00 | 7.90 | 8.30 |
| Chile | CHL | 11.00 | 9.40 | 8.50 | 8.20 | | 6.10 |
| China | CHN | 3.10 | 3.10 | 3.70 | 4.30 | 4.00 | 3.60 |
| Colombia | COL | 5.40 | 5.30 | 5.20 | 1.30 | 1.20 | 8.40 |
| Croacia | HRV | | 31.80 | 29.70 | 23.90 | 24.10 | 27.20 |
| Cuba | CUB | 9.30 | 8.40 | 8.10 | 7.10 | 8.30 | 8.20 |
| Chipre | CYP | 10.10 | 9.00 | 6.90 | 5.60 | 6.80 | 10.20 |
| República Checa | CZE | 2.70 | 2.30 | 3.30 | 14.30 | 14.10 | 6.10 |
| Dinamarca | DNK | 6.80 | 6.30 | 6.00 | 6.10 | 8.30 | 8.00 |
| Ecuador | ECU | 8.00 | 7.70 | 7.10 | 7.80 | 9.70 | 8.10 |
| Egipto, República Árabe | EGY | 4.20 | 4.10 | 4.00 | | 4.30 | 14.20 |
| Estonia | EST | 11.30 | 10.50 | 12.00 | 13.20 | 12.00 | 11.60 |
| Finlandia | FIN | 6.60 | 6.00 | 4.60 | 4.90 | 7.80 | 10.60 |
| Francia | FRA | 12.60 | 11.10 | 9.60 | 8.40 | 9.00 | 11.80 |
| Georgia | GEO | 1.90 | 1.90 | 1.80 | 1.60 | 7.40 | 8.00 |
| Alemania | DEU | 5.30 | 4.50 | 3.90 | 3.60 | 5.30 | 6.20 |
| Grecia | GRC | 7.90 | 7.10 | 5.30 | 4.40 | 6.70 | 7.30 |
| Guatemala | GTM | 4.80 | 3.90 | 3.80 | 3.30 | 6.00 | 7.40 |
| Hong Kong, Región Administrativa | HKG | 18.00 | 16.40 | 15.60 | 14.20 | 14.90 | 14.30 |
| Hungría | HUN | 7.70 | 7.70 | 6.10 | 7.30 | 6.50 | 7.20 |
| Islandia | ISL | 11.20 | 10.60 | 8.90 | 8.70 | 9.40 | 3.10 |
| India | IND | 7.20 | 6.60 | 6.30 | 5.90 | 7.30 | 4.60 |
| Irlanda | IRL | 7.90 | 5.90 | 4.70 | 5.50 | 13.80 | 16.90 |
| Israel | ISR | 8.40 | 7.60 | 6.80 | 6.30 | 8.20 | 8.40 |
| Japón | JPN | 8.90 | 8.80 | 8.00 | 7.40 | 9.10 | 9.30 |
| Jordania | JOR | 13.80 | 13.60 | 13.30 | 16.50 | | 5.10 |
| Kenya | KEN | 11.10 | 10.30 | 8.60 | 7.50 | 7.70 | 7.10 |
| Corea, Rep. Popular Democrática | PRK | 9.90 | 8.90 | 8.30 | 7.70 | 9.50 | 12.50 |
| Corea, República de | KOR | 5.60 | 4.80 | 4.00 | 3.60 | 5.20 | 12.20 |
| Letonia | LVA | 7.20 | 7.50 | 7.40 | 7.80 | 10.00 | 11.20 |
| Lituania | LTU | 2.60 | 3.00 | 2.30 | 3.00 | 7.20 | 7.60 |
| Luxemburgo | LUX | 11.20 | 10.30 | 9.10 | 8.40 | 7.90 | 7.10 |
| Madagascar | MDG | 12.10 | | 10.60 | 10.50 | | 10.60 |
| Malasia | MYS | 4.30 | 4.40 | 4.60 | 6.00 | 11.70 | 13.50 |
| México | MEX | 9.00 | 8.40 | 7.30 | 6.10 | 7.50 | 6.60 |
| República de Moldova | MDA | 7.70 | 6.80 | 6.10 | 6.70 | 7.80 | 8.40 |
| Mónaco | MCO | 10.90 | 9.60 | 9.40 | 10.60 | 11.40 | 6.80 |
| Mongolia | MNG | 4.40 | 4.10 | 3.90 | 4.00 | 5.00 | 5.00 |
| Marruecos | MAR | 8.10 | 7.80 | 7.30 | 6.60 | 6.60 | 5.40 |
| Países Bajos | NLD | 3.70 | 3.40 | 3.20 | 3.20 | 3.60 | 3.70 |
| Nueva Zelandia | NZL | 41.40 | 44.90 | 46.30 | 47.50 | 45.40 | 4.60 |
| Noruega | NOR | 8.10 | 8.30 | 8.20 | 8.20 | 8.40 | 8.60 |
| Pakistán | PAK | 8.90 | 6.80 | 6.00 | 7.40 | 17.10 | 18.70 |
| Paraguay | PRY | 8.30 | 5.60 | 4.30 | 5.80 | 13.70 | 17.80 |
| Perú | PER | 4.50 | 4.70 | 4.10 | 5.10 | 5.10 | 4.40 |
| Filipinas | PHL | 4.20 | 3.70 | 3.00 | 3.00 | 3.60 | 6.40 |
| Polonia | POL | 37.30 | 36.00 | 34.90 | 33.80 | 32.20 | 32.00 |
| Portugal | PRT | 3.50 | 3.30 | 3.20 | 3.30 | 3.70 | 5.20 |
| Rumania | ROU | 7.30 | 6.90 | 6.50 | 6.00 | 6.90 | 6.90 |
| Federación de Rusia | RUS | 9.60 | 9.10 | 8.50 | 7.20 | 7.30 | 7.70 |
| Serbia | SRB | 9.30 | 9.60 | 7.60 | 7.30 | 7.70 | 6.40 |
| Singapur | SGP | 3.50 | 3.20 | 3.40 | 3.50 | 5.20 | 5.30 |
| República Eslovaca | SVK | 7.30 | 7.40 | 5.10 | 4.00 | 6.40 | 4.60 |
| Eslovenia | SVN | 11.00 | 9.70 | 9.70 | 9.60 | 10.00 | 9.20 |
| Sudáfrica | ZAF | 4.70 | 3.90 | 3.20 | 2.80 | 3.40 | 4.50 |
| España | ESP | 3.80 | 3.90 | 3.70 | 4.20 | 6.10 | 6.50 |
| Sri Lanka | LKA | 4.60 | 3.40 | 2.50 | 2.60 | 3.20 | 3.60 |
| Suecia | SWE | 9.80 | 8.70 | 6.40 | 5.60 | 6.60 | 6.50 |
| Suiza | CHE | 5.80 | 6.70 | 5.60 | 5.60 | 8.80 | 5.30 |
| Tailandia | THA | 11.40 | 8.80 | 7.20 | 6.40 | 6.30 | 5.80 |
| Ucrania | UKR | 7.70 | 8.00 | 7.40 | 7.30 | 7.50 | 7.40 |
| Reino Unido | GBR | 17.70 | 13.80 | 9.60 | 7.10 | 8.20 | 9.60 |
| Estados Unidos | USA | 7.60 | 7.70 | 8.00 | 7.60 | 9.50 | 10.80 |
| Uruguay | URY | 11.30 | 11.00 | 10.90 | 11.60 | 13.40 | 5.40 |
| Uzbekistán | UZB | 7.20 | 7.30 | 6.40 | 5.80 | 6.90 | 7.30 |
| Viet Nam | VNM | 7.20 | 7.20 | 6.10 | 6.30 | 8.40 | 7.50 |

Anexo 2
Nivel de empleo en el mundo según el Banco Mundial

| Country Name | Code | 2005 | 2006 | 2007 | 2008 | 2009 | 2010 |
|---|---|---|---|---|---|---|---|
| Asia oriental y el Pacífico | EAS | 21.40 | 20.10 | 19.20 | 19.10 | 19.60 | 20.10 |
| Asia oriental y el Pacífico | EAP | 9.80 | 9.50 | 9.30 | 9.00 | 9.00 | 8.60 |
| Zona del Euro | EMU | 8.50 | 8.50 | 9.20 | 9.00 | 8.70 | 9.00 |
| Europa y Asia central | ECS | 62.60 | 56.50 | 53.20 | 54.70 | 53.30 | 54.10 |
| Europa y Asia central | ECA | 2.20 | 2.20 | 2.10 | 2.00 | 2.00 | 2.00 |
| Unión Europea | EUU | 10.60 | 10.40 | 10.30 | 10.00 | 10.30 | 9.90 |
| Ingreso alto | HIC | 28.40 | 27.20 | 27.00 | 25.20 | 25.10 | |
| Ingreso alto: Miembros de OCDE | OEC | 10.20 | 9.20 | 8.40 | 8.70 | 9.00 | 9.00 |
| América Latina y el Caribe | LCN | 27.00 | 26.00 | 24.80 | 24.40 | 25.70 | 24.70 |
| América Latina y el Caribe | LAC | 45.40 | 43.70 | 41.30 | 46.90 | 47.20 | 48.60 |
| Ingreso mediano y bajo | LMY | 21.10 | 21.60 | 19.70 | 19.60 | 10.10 | 10.30 |
| Países de ingreso mediano bajo | LMC | 19.90 | 17.90 | 16.20 | 16.30 | 16.60 | 17.80 |
| Ingreso mediano | MIC | 15.80 | 15.10 | 14.30 | 14.40 | 14.70 | 13.90 |
| América del Norte | NAC | 12.30 | 12.10 | 12.40 | 12.50 | 13.00 | 14.10 |
| Miembros OCDE | OED | 5.20 | 5.10 | 5.10 | 5.00 | 5.30 | 5.30 |
| Asia meridional | SAS | 13.30 | 12.70 | 10.80 | 12.20 | 11.30 | 12.90 |
| Ingreso mediano alto | UMC | 41.60 | 42.50 | 41.90 | 40.00 | 42.50 | 45.60 |
| Mundo | WLD | 25.30 | 24.80 | 27.30 | 22.70 | 21.80 | 20.40 |
| Argentina | ARG | 38.80 | 35.50 | 35.10 | 36.20 | 38.30 | |
| Armenia | ARM | 5.40 | 5.40 | 5.80 | 4.40 | 4.40 | 4.80 |
| Australia | AUS | 8.80 | 8.80 | 8.60 | 9.00 | 9.60 | 9.20 |
| Austria | AUT | 6.60 | 6.90 | 6.50 | 5.90 | 6.50 | 7.00 |
| Belarús | BLR | 64.20 | 64.60 | 62.20 | 63.20 | 61.20 | 61.90 |
| Bélgica | BEL | 7.40 | 7.20 | 7.10 | 6.80 | 6.70 | 6.70 |
| Brasil | BRA | 28.40 | 28.30 | 27.60 | 27.00 | 27.30 | 28.00 |
| Bulgaria | BGR | 47.70 | 49.90 | 50.10 | 53.40 | 51.60 | 52.80 |
| Canadá | CAN | 7.70 | 7.80 | 7.20 | 7.40 | 7.40 | |
| Chile | CHL | 7.70 | 7.10 | 7.00 | 7.00 | 7.10 | 6.80 |
| China | CHN | 8.70 | 9.40 | 8.70 | 8.00 | 8.20 | 8.30 |
| Colombia | COL | 63.40 | 63.10 | 63.10 | 64.40 | 63.70 | 61.80 |
| Croacia | HRV | 11.40 | 10.90 | 11.40 | 11.70 | 12.00 | 11.80 |
| Cuba | CUB | 12.60 | 12.30 | 12.10 | 12.10 | 11.90 | 12.10 |
| Chipre | CYP | 7.50 | 7.90 | 7.40 | 7.20 | | 16.80 |
| República Checa | CZE | 19.50 | 19.30 | 19.10 | 18.60 | 18.20 | 18.50 |
| Dinamarca | DNK | 34.50 | 35.40 | 33.40 | 36.50 | 31.90 | 82.90 |
| Ecuador | ECU | 12.10 | 11.20 | 10.80 | 10.50 | 10.10 | 9.30 |
| Egipto, República Árabe | EGY | 3.00 | 2.80 | 0.90 | 2.10 | 9.60 | 29.50 |
| Estonia | EST | 34.40 | 33.60 | 32.80 | 31.90 | 61.90 | 32.40 |
| Finlandia | FIN | 26.30 | 25.80 | 25.20 | 24.80 | 23.50 | 51.10 |
| Francia | FRA | 8.20 | 8.20 | 7.40 | 6.80 | 7.70 | 7.50 |
| Georgia | GEO | 14.90 | 13.50 | 11.60 | 9.40 | 9.60 | 8.80 |
| Alemania | DEU | 5.30 | 5.20 | 4.30 | 4.10 | 5.90 | 4.60 |
| Grecia | GRC | 6.40 | 5.50 | 5.20 | 4.60 | 8.20 | 7.80 |
| Guatemala | GTM | 22.40 | 23.30 | 22.20 | 23.20 | 8.40 | 12.10 |
| Hong Kong, Región Administrativa | HKG | 21.20 | 21.90 | 22.30 | 21.90 | 21.50 | 45.20 |
| Hungría | HUN | 9.20 | 9.50 | 9.30 | 9.00 | 9.30 | 9.70 |
| Islandia | ISL | 16.70 | 16.20 | 16.80 | 16.30 | 15.90 | 16.00 |
| India | IND | 31.00 | 29.70 | 29.50 | 29.20 | 29.50 | 16.30 |
| Irlanda | IRL | 36.30 | 31.90 | 32.40 | 31.10 | 28.50 | 29.20 |
| Israel | ISR | 35.40 | 36.30 | 61.90 | 53.10 | 57.50 | 31.10 |
| Japón | JPN | 57.70 | 51.90 | 51.10 | 51.90 | 50.50 | 53.10 |
| Jordania | JOR | 8.60 | 8.90 | 9.20 | 9.40 | 9.70 | 11.10 |
| Kenya | KEN | 11.90 | 12.10 | 11.80 | 12.10 | 10.90 | 51.90 |
| Corea, Rep. Popular Democrática | PRK | 45.40 | 45.20 | 44.70 | 32.70 | 32.40 | 59.70 |
| Corea, República de | KOR | 6.10 | 6.40 | 5.90 | 5.70 | 5.90 | 5.60 |
| Letonia | LVA | 61.20 | 61.90 | 61.80 | 63.10 | 35.40 | 55.80 |
| Lituania | LTU | 31.80 | 30.50 | 27.70 | 27.20 | 32.20 | 30.90 |
| Luxemburgo | LUX | 48.80 | 49.20 | 46.80 | 44.50 | 87.80 | 16.80 |
| Madagascar | MDG | 33.20 | 32.70 | 31.20 | 39.60 | 82.90 | 61.90 |
| Malasia | MYS | 44.80 | 44.50 | 43.60 | 43.50 | 9.30 | 51.10 |
| México | MEX | 21.80 | 20.40 | 19.40 | 18.90 | 18.60 | 18.60 |
| República de Moldova | MDA | 19.70 | 19.00 | 19.00 | 18.50 | 18.30 | 17.60 |
| Mónaco | MCO | 33.50 | 32.10 | 32.20 | 31.20 | 31.40 | 33.10 |
| Mongolia | MNG | 6.20 | 6.00 | 5.80 | 5.70 | 29.50 | 32.40 |
| Marruecos | MAR | 24.40 | 22.50 | 22.70 | 29.40 | 28.20 | 5.20 |
| Países Bajos | NLD | 10.30 | 10.00 | 10.30 | 10.00 | 9.80 | 5.50 |
| Nueva Zelandia | NZL | 9.40 | 9.50 | 9.80 | 10.60 | 12.30 | 12.40 |
| Noruega | NOR | 11.30 | 15.90 | 13.00 | 11.50 | 12.80 | 13.50 |
| Pakistán | PAK | 17.60 | 15.60 | 15.10 | 10.10 | 10.10 | 16.80 |
| Paraguay | PRY | 13.00 | 12.40 | 12.10 | 11.80 | 11.20 | 11.20 |
| Perú | PER | 37.60 | 41.30 | 40.70 | 40.60 | 39.80 | 23.30 |
| Filipinas | PHL | 6.60 | 6.80 | 6.70 | 6.60 | 6.90 | 7.00 |
| Polonia | POL | 9.60 | 9.70 | 10.10 | 9.70 | 9.40 | 9.10 |
| Portugal | PRT | 44.70 | 34.90 | 37.50 | 31.50 | 34.20 | 33.10 |
| Rumania | ROU | 53.10 | 53.20 | 53.30 | 54.10 | 52.50 | 53.10 |
| Federación de Rusia | RUS | 40.70 | 38.10 | 33.80 | 33.10 | 34.30 | 33.80 |
| Serbia | SRB | 10.10 | 10.30 | 10.50 | 10.50 | 10.80 | 11.30 |
| Singapur | SGP | 24.70 | 25.10 | 25.10 | 23.40 | 23.20 | 42.40 |
| República Eslovaca | SVK | 31.20 | 30.40 | 29.80 | 30.50 | 30.80 | 53.20 |
| Eslovenia | SVN | 36.20 | 36.10 | 36.20 | 30.40 | 27.60 | 92.20 |
| Sudáfrica | ZAF | 36.90 | 34.30 | 34.40 | 35.30 | 33.10 | 56.70 |
| España | ESP | 80.10 | 77.50 | 78.30 | 77.00 | 73.90 | 17.50 |
| Sri Lanka | LKA | 37.60 | 36.20 | 36.10 | 36.20 | 30.40 | 47.50 |
| Suecia | SWE | 40.70 | 38.10 | 33.80 | 33.10 | 34.30 | 9.00 |
| Suiza | CHE | 12.40 | 12.10 | 11.80 | 11.20 | 11.20 | 24.90 |
| Tailandia | THA | 41.30 | 40.70 | 40.60 | 39.80 | | 33.20 |
| Ucrania | UKR | 6.80 | 6.70 | 6.60 | 6.90 | 7.00 | 56.10 |
| Reino Unido | GBR | 9.70 | 10.10 | 9.70 | 9.40 | 9.10 | 16.50 |
| Estados Unidos | USA | 34.90 | 37.50 | 31.50 | 34.20 | 33.10 | 44.40 |
| Uruguay | URY | 7.00 | 7.00 | 6.60 | 6.90 | 6.60 | 9.80 |
| Uzbekistán | UZB | 10.40 | 10.40 | 10.90 | 9.80 | 9.60 | 26.40 |
| Viet Nam | VNM | 39.10 | 39.30 | 39.50 | 37.60 | 37.60 | 39.20 |

## Anexo 3
## Nivel de tecnología en el mundo según el Banco Mundial

| Country Name | Code | 2005 | 2006 | 2007 | 2008 | 2009 | 2010 |
|---|---|---|---|---|---|---|---|
| Asia oriental y el Pacífico | EAS | 1.44 | 1.35 | 1.09 | 1.46 | 1.44 | 1.84 |
| Asia oriental y el Pacífico | EAP | 2.76 | 2.55 | 2.10 | 1.14 | 2.41 | 2.88 |
| Zona del Euro | EMU | 30.71 | 30.27 | 27.01 | 25.65 | 27.35 | 26.37 |
| Europa y Asia central | ECS | 32.87 | 32.13 | 29.01 | 26.33 | 28.41 | 28.72 |
| Europa y Asia central | ECA | 17.13 | 16.68 | 13.69 | 13.21 | 14.77 | 14.93 |
| Unión Europea | EUU | 17.27 | 17.46 | 13.59 | 13.12 | 14.99 | 14.77 |
| Ingreso alto | HIC | 6.07 | 6.64 | 6.19 | 6.31 | 8.22 | 6.70 |
| Ingreso alto: Miembros de OCDE | OEC | 18.16 | 18.41 | 14.02 | 13.57 | 15.31 | 15.33 |
| América Latina y el Caribe | LCN | 5.27 | 6.23 | 4.80 | 3.41 | 7.19 | 20.86 |
| América Latina y el Caribe | LAC | 21.32 | 21.51 | 17.59 | 17.08 | 18.44 | 17.40 |
| Ingreso mediano y bajo | LMY | 24.94 | 25.75 | 20.28 | 20.14 | 25.45 | 24.80 |
| Países de ingreso mediano bajo | LMC | 20.75 | 20.86 | 17.15 | 16.56 | 17.41 | 16.50 |
| Ingreso mediano | MIC | 12.17 | 12.02 | 11.61 | 10.05 | 11.47 | 10.93 |
| América del Norte | NAC | 12.39 | 12.23 | 11.85 | 10.27 | 11.52 | 10.93 |
| Miembros OCDE | OED | 18.64 | 18.55 | 17.38 | 15.57 | 17.88 | 17.78 |
| Asia meridional | SAS | 12.72 | 11.72 | 10.79 | 9.35 | 11.39 | 11.03 |
| Ingreso mediano alto | UMC | 3.04 | 3.59 | 2.02 | 2.52 | 3.49 | 32.48 |
| Mundo | WLD | 18.84 | 18.72 | 17.54 | 15.67 | 18.01 | 17.90 |
| Argentina | ARG | 25.74 | 26.06 | 23.89 | 23.18 | 20.46 | 18.75 |
| Armenia | ARM | 20.40 | 20.48 | 16.88 | 16.26 | 17.12 | 16.22 |
| Australia | AUS | 3.54 | 2.75 | 3.80 | 2.78 | 3.90 | 0.72 |
| Austria | AUT | 4.09 | 7.76 | 4.62 | 4.25 | 0.78 | 1.08 |
| Belarús | BLR | 4.98 | 5.25 | 5.67 | 6.29 | 8.28 | 6.67 |
| Bélgica | BEL | 3.10 | 7.34 | 4.43 | 3.60 | 5.05 | 2.80 |
| Brasil | BRA | 54.65 | 7.34 | 4.43 | 3.60 | 5.05 | 2.80 |
| Bulgaria | BGR | 20.17 | 20.36 | 19.17 | 17.16 | 19.67 | 19.45 |
| Canadá | CAN | 20.67 | 20.74 | 17.54 | 16.67 | 18.29 | 17.52 |
| Chile | CHL | 4.22 | 6.17 | 1.31 | 3.47 | 0.76 | 0.89 |
| China | CHN | 1.48 | 1.62 | 0.72 | 0.66 | 0.63 | 0.50 |
| Colombia | COL | 6.83 | 7.05 | 6.59 | 9.02 | 8.69 | 7.45 |
| Croacia | HRV | 0.62 | 1.08 | 0.88 | 1.41 | 2.18 | 1.85 |
| Cuba | CUB | 6.41 | 19.64 | 18.98 | 17.18 | 3.03 | 3.29 |
| Chipre | CYP | 12.79 | 12.34 | 10.27 | 10.79 | 11.93 | 11.88 |
| República Checa | CZE | 13.74 | 13.34 | 11.31 | 11.00 | 11.64 | 11.91 |
| Dinamarca | DNK | 1.06 | 1.75 | 3.40 | 0.92 | 0.98 | 1.08 |
| Ecuador | ECU | 0.00 | 0.39 | 0.00 | 9.64 | 9.79 | 8.82 |
| Egipto, República Árabe | EGY | 0.07 | 0.06 | 0.05 | 8.37 | 2.31 | 10.57 |
| Estonia | EST | 0.32 | 0.22 | 1.15 | 11.95 | 6.50 | 5.19 |
| Finlandia | FIN | 20.35 | 18.06 | 9.99 | 4.27 | 9.80 | 12.13 |
| Francia | FRA | 2.70 | 2.79 | 2.77 | 2.43 | 3.13 | 3.04 |
| Georgia | GEO | 9.02 | 8.35 | 7.41 | 7.98 | 10.36 | 10.47 |
| Alemania | DEU | 0.01 | 10.21 | 0.08 | 0.11 | 0.38 | 30.89 |
| Grecia | GRC | 4.14 | 5.49 | 0.75 | 0.13 | 0.22 | 0.14 |
| Guatemala | GTM | 8.94 | 22.30 | 4.68 | 4.43 | 4.93 | 8.62 |
| Hong Kong, Región Administrativa | HKG | 2.04 | 2.17 | 2.14 | 4.14 | 3.28 | 2.58 |
| Hungría | HUN | 0.44 | 0.42 | 0.43 | 0.57 | 0.92 | 0.40 |
| Islandia | ISL | 12.84 | 12.08 | 11.87 | 11.65 | 13.20 | 11.21 |
| India | IND | 4.77 | 6.12 | 5.97 | 6.55 | 8.15 | 7.91 |
| Irlanda | IRL | 22.48 | 11.16 | 5.67 | 0.68 | 0.51 | 7.80 |
| Israel | ISR | 4.19 | 3.14 | 2.16 | 7.52 | 11.54 | 8.50 |
| Japón | JPN | 0.12 | 0.03 | 0.04 | 0.08 | 0.18 | 0.09 |
| Jordania | JOR | 2.21 | 2.74 | 2.93 | 1.97 | 2.54 | 4.87 |
| Kenya | KEN | 13.08 | 13.34 | 12.75 | 13.60 | 16.22 | 14.04 |
| Corea, Rep. Popular Democrática | PRK | 6.42 | 6.38 | 6.48 | 4.89 | 3.40 | 5.48 |
| Corea, República de | KOR | 30.84 | 30.51 | 26.66 | 25.57 | 27.53 | 27.51 |
| Letonia | LVA | 4.99 | 4.08 | 2.91 | 3.72 | 5.22 | 5.06 |
| Lituania | LTU | 38.03 | 44.72 | 45.37 | 39.43 | 44.18 | 39.97 |
| Luxemburgo | LUX | 11.40 | 9.85 | 8.21 | 8.35 | 9.76 | 9.15 |
| Madagascar | MDG | 16.97 | 22.73 | 29.37 | 30.45 | 30.86 | 37.12 |
| Malasia | MYS | 12.95 | 14.32 | 13.24 | 13.56 | 14.56 | 15.30 |
| México | MEX | 22.51 | 19.93 | 16.79 | 15.58 | 17.70 | 14.21 |
| República de Moldova | MDA | 7.48 | 7.74 | 6.18 | 0.45 | 0.93 | 0.01 |
| Mónaco | MCO | 8.40 | 5.58 | 13.67 | 5.33 | 2.75 | 2.95 |
| Mongolia | MNG | 7.66 | 7.87 | 6.17 | 3.72 | 3.93 | 8.43 |
| Marruecos | MAR | 0.40 | 0.55 | 0.19 | 0.97 | 0.84 | 0.88 |
| Países Bajos | NLD | 4.34 | 4.42 | 4.53 | 4.27 | 4.99 | 5.78 |
| Nueva Zelandia | NZL | 14.66 | 12.63 | 5.80 | 5.40 | 5.68 | 9.15 |
| Noruega | NOR | 1.45 | 0.43 | 2.53 | 2.62 | 3.09 | 3.03 |
| Pakistán | PAK | 6.15 | 8.06 | 10.85 | 11.14 | 9.99 | 10.61 |
| Paraguay | PRY | 11.86 | 11.58 | 8.76 | 6.42 | 8.78 | 8.37 |
| Perú | PER | 25.06 | 22.31 | 17.98 | 17.21 | 13.96 | 10.80 |
| Filipinas | PHL | 20.27 | 21.46 | 18.48 | 19.97 | 22.64 | 24.92 |
| Polonia | POL | 13.15 | 11.68 | 2.33 | 16.02 | 9.11 | 4.75 |
| Portugal | PRT | 5.31 | 6.82 | 6.95 | 6.95 | 7.76 | 7.64 |
| Rumania | ROU | 2.83 | 2.76 | 2.24 | 2.44 | 4.53 | 12.80 |
| Federación de Rusia | RUS | 22.59 | 16.47 | 7.49 | 2.69 | 3.94 | 1.78 |
| Serbia | SRB | 17.42 | 17.14 | 13.99 | 13.30 | 15.26 | 15.25 |
| Singapur | SGP | 0.20 | 0.18 | 1.05 | 1.43 | 3.66 | 2.00 |
| República Eslovaca | SVK | 10.58 | 10.96 | 7.37 | 9.31 | 10.86 | 10.19 |
| Eslovenia | SVN | 3.38 | 3.44 | 3.56 | 4.20 | 4.59 | 5.68 |
| Sudáfrica | ZAF | 22.98 | 22.06 | 18.41 | 17.31 | 18.76 | 17.96 |
| España | ESP | 1.39 | 1.23 | 1.13 | 0.92 | 1.41 | 2.86 |
| Sri Lanka | LKA | 2.90 | 3.24 | 5.51 | 4.15 | 5.31 | 5.70 |
| Suecia | SWE | 15.60 | 11.36 | 11.52 | 11.25 | 14.71 | 16.10 |
| Suiza | CHE | 25.83 | 24.12 | 23.79 | 23.30 | 24.94 | 24.24 |
| Tailandia | THA | 33.97 | 46.88 | 60.66 | 40.75 | 31.45 | 20.87 |
| Ucrania | UKR | 5.80 | 6.07 | 6.40 | 6.78 | 9.09 | 7.18 |
| Reino Unido | GBR | 16.55 | 13.47 | 11.00 | 10.90 | 12.87 | 11.37 |
| Estados Unidos | USA | 34.73 | 34.53 | 27.26 | 25.73 | 24.26 | 21.23 |
| Uruguay | URY | 14.03 | 14.51 | 7.48 | 11.12 | 17.62 | 14.66 |
| Uzbekistán | UZB | 7.98 | 7.33 | 6.26 | 6.40 | 7.47 | 7.23 |
| Viet Nam | VNM | 0.09 | 0.11 | 0.20 | 0.43 | 0.51 | 0.57 |

## Anexo 4
## Nivel de PIB por empleo en el mundo según el Banco Mundial

| Country Name | Code | 2005 | 2006 | 2007 | 2008 | 2009 | 2010 |
|---|---|---|---|---|---|---|---|
| Asia oriental y el Pacífico | EAS | 11338.72 | 12104.76 | 13056.00 | 13647.27 | 14295.89 | 15265.54 |
| Asia oriental y el Pacífico | EAP | 7999.17 | 8770.28 | 9721.89 | 10395.90 | 11195.23 | 12096.46 |
| Zona del Euro | EMU | 44606.21 | 45200.62 | 45690.42 | 45559.65 | 44554.39 | 45512.88 |
| Europa y Asia central | ECS | 30912.77 | 31788.43 | 32531.62 | 32693.91 | 31686.05 | 32367.15 |
| Europa y Asia central | ECA | 15464.68 | 16561.78 | 17623.07 | 18205.49 | 17329.92 | 17947.95 |
| Unión Europea | EUU | 41077.64 | 41724.50 | 42256.22 | 42130.15 | 41105.71 | 42032.91 |
| Ingreso alto | HIC | 2233.73 | 2321.62 | 2412.90 | 2507.94 | 2558.58 | 2609.99 |
| Ingreso alto: Miembros de OCDE | OEC | 48883.76 | 49485.19 | 49977.66 | 49844.13 | 49186.14 | 50547.46 |
| América Latina y el Caribe | LCN | 49706.31 | 50342.78 | 50890.04 | 50804.45 | 50233.27 | 51661.98 |
| América Latina y el Caribe | LAC | 16239.54 | 16647.45 | 17301.37 | 17572.64 | 17081.20 | 17607.48 |
| Ingreso mediano y bajo | LMY | 16158.27 | 16551.43 | 17202.15 | 17474.29 | 16982.04 | 17509.41 |
| Países de ingreso mediano bajo | LMC | 2729.19 | 2869.36 | 3011.58 | 3111.13 | 2938.74 | 3029.99 |
| Ingreso mediano | MIC | 8484.26 | 9053.84 | 9748.63 | 10201.16 | 10498.79 | 11098.06 |
| América del Norte | NAC | 2555.91 | 2682.82 | 2812.76 | 2895.53 | 2692.97 | 2764.57 |
| Miembros OCDE | OED | 6934.59 | 7294.99 | 7702.54 | 7976.91 | 8213.15 | 8582.24 |
| Asia meridional | SAS | 14350.87 | 14762.88 | 15173.94 | 15506.46 | 15493.79 | 15725.21 |
| Ingreso mediano alto | UMC | 11832.35 | 12177.49 | 12586.05 | 12798.27 | 12902.78 | 13090.45 |
| Mundo | WLD | 9175.10 | 9803.82 | 10572.62 | 11082.52 | 11359.73 | 12030.06 |
| Argentina | ARG | 62270.92 | 62758.94 | 63206.90 | 63385.51 | 63960.94 | 66022.14 |
| Armenia | ARM | 46200.19 | 46761.90 | 47293.20 | 47139.29 | 46454.22 | 47627.19 |
| Australia | AUS | 6200.03 | 6561.68 | 6970.85 | 7274.29 | 7607.98 | 7985.77 |
| Austria | AUT | 3103.31 | 3226.61 | 3370.97 | 3456.05 | 3481.75 | 3558.13 |
| Belarús | BLR | 3103.31 | 3226.61 | 3370.97 | 3456.05 | 3481.75 | 3558.13 |
| Bélgica | BEL | 10831.23 | 11663.29 | 12698.21 | 13398.84 | 13721.47 | 14613.07 |
| Brasil | BRA | 15714.27 | 16297.52 | 16957.60 | 17286.16 | 17323.45 | 17980.03 |
| Bulgaria | BGR | 13384.00 | 14065.00 | 14896.00 | 15897.00 | 17787.00 | 18078.00 |
| Canadá | CAN | 8372.00 | 8256.00 | 8245.00 | 8210.00 | 8214.00 | 8334.00 |
| Chile | CHL | 1825.00 | 2101.00 | 2454.00 | 2698.00 | 2653.00 | 2744.00 |
| China | CHN | 24767.00 | 25789.00 | 27746.00 | 28581.00 | 27478.00 | 28678.00 |
| Colombia | COL | 22872.00 | 26018.00 | 29351.00 | 30938.00 | 26270.00 | 27029.00 |
| Croacia | HRV | 48482.00 | 48928.00 | 49463.00 | 49427.00 | 49677.00 | 50153.00 |
| Cuba | CUB | 46226.00 | 47226.00 | 48119.00 | 48317.00 | 46863.00 | 47474.00 |
| Chipre | CYP | 9620.00 | 12535.00 | 15514.00 | 17005.00 | 18509.00 | 18939.00 |
| República Checa | CZE | 12209.00 | 12840.00 | 13702.00 | 14378.00 | 14631.00 | 14940.00 |
| Dinamarca | DNK | 3245.00 | 3385.00 | 3524.00 | 3662.00 | 3787.00 | 3917.00 |
| Ecuador | ECU | 19407.00 | 19877.00 | 20384.00 | 20289.00 | 19463.00 | 19259.00 |
| Egipto, República Árabe | EGY | 21432.00 | 23296.00 | 24716.00 | 26720.00 | 26598.00 | 28465.00 |
| Estonia | EST | 54235.00 | 55036.00 | 55752.00 | 55351.00 | 54022.00 | 54882.00 |
| Finlandia | FIN | 7250.00 | 7398.00 | 7510.00 | 7762.00 | 7842.00 | 7964.00 |
| Francia | FRA | 25874.00 | 27363.00 | 27367.00 | 27789.00 | 26799.00 | 26751.00 |
| Georgia | GEO | 12059.00 | 12239.00 | 12732.00 | 12947.00 | 12891.00 | 13419.00 |
| Alemania | DEU | 16014.00 | 16505.00 | 17029.00 | 17617.00 | 17207.00 | 18141.00 |
| Grecia | GRC | 2455.00 | 2506.00 | 2513.00 | 2566.00 | 2559.00 | 2581.00 |
| Guatemala | GTM | 3343.00 | 3615.00 | 3883.00 | 4089.00 | 3906.00 | 3988.00 |
| Hong Kong, Región Administrativa | HKG | 2862.00 | 2872.00 | 2885.00 | 2888.00 | 2871.00 | 2872.00 |
| Hungría | HUN | 48862.00 | 49280.00 | 49225.00 | 48729.00 | 48302.00 | 48916.00 |
| Islandia | ISL | 30473.00 | 29356.00 | 30004.00 | 30153.00 | 29259.00 | 30417.00 |
| India | IND | 7710.00 | 8620.00 | 9768.00 | 10638.00 | 11540.00 | 12593.00 |
| Irlanda | IRL | 15326.00 | 16856.00 | 17708.00 | 17775.00 | 17501.00 | 17985.00 |
| Israel | ISR | 633.00 | 647.00 | 666.00 | 684.00 | 679.00 | 690.00 |
| Japón | JPN | 16505.00 | 17434.00 | 17858.00 | 18025.00 | 17464.00 | 17777.00 |
| Jordania | JOR | 3008.00 | 2954.00 | 2926.00 | 2916.00 | 2947.00 | 2955.00 |
| Kenya | KEN | 22332.00 | 23529.00 | 23987.00 | 24292.00 | 23310.00 | 24003.00 |
| Corea, Rep. Popular Democrática | PRK | 25701.00 | 26302.00 | 26782.00 | 26988.00 | 26743.00 | 27080.00 |
| Corea, República de | KOR | 22641.00 | 23723.00 | 24524.00 | 24822.00 | 24049.00 | 24941.00 |
| Letonia | LVA | 47151.00 | 47751.00 | 47204.00 | 45822.00 | 44814.00 | 46598.00 |
| Lituania | LTU | 11440.00 | 11964.00 | 12682.00 | 12972.00 | 13038.00 | 13487.00 |
| Luxemburgo | LUX | 12159.00 | 12193.00 | 12494.00 | 13208.00 | 12883.00 | 12982.00 |
| Madagascar | MDG | 10641.00 | 11070.00 | 11664.00 | 12240.00 | 12530.00 | 12897.00 |
| Malasia | MYS | 40686.00 | 42687.00 | 45299.00 | 42932.00 | 41032.00 | 44568.00 |
| México | MEX | 1543.00 | 1668.00 | 1810.00 | 1952.00 | 2083.00 | 2184.00 |
| República de Moldova | MDA | 49113.00 | 50352.00 | 51909.00 | 51574.00 | 48817.00 | 50278.00 |
| Mónaco | MCO | 53990.00 | 54650.00 | 55149.00 | 54941.00 | 54171.00 | 55033.00 |
| Mongolia | MNG | 12661.00 | 13828.00 | 15926.00 | 17336.00 | 16589.00 | 17433.00 |
| Marruecos | MAR | 42411.00 | 43570.00 | 43998.00 | 43823.00 | 41760.00 | 43050.00 |
| Países Bajos | NLD | 3205.00 | 3321.00 | 3423.00 | 3577.00 | 3629.00 | 3711.00 |
| Nueva Zelandia | NZL | 34254.00 | 34647.00 | 35503.00 | 35881.00 | 35311.00 | 34767.00 |
| Noruega | NOR | 13150.00 | 13401.00 | 13735.00 | 14288.00 | 13700.00 | 13588.00 |
| Pakistán | PAK | 53841.00 | 56522.00 | 58707.00 | 59402.00 | 58386.00 | 61382.00 |
| Paraguay | PRY | 20737.00 | 21358.00 | 21588.00 | 22049.00 | 21173.00 | 21473.00 |
| Perú | PER | 45204.00 | 44969.00 | 45602.00 | 45677.00 | 45462.00 | 44155.00 |
| Filipinas | PHL | 6283.00 | 6727.00 | 7169.00 | 7528.00 | 7959.00 | 8401.00 |
| Polonia | POL | 9140.00 | 9492.00 | 9843.00 | 9933.00 | 10193.00 | 10587.00 |
| Portugal | PRT | 14057.00 | 14651.00 | 15573.00 | 15693.00 | 15515.00 | 15415.00 |
| Rumania | ROU | 5553.00 | 5736.00 | 5664.00 | 6045.00 | 6109.00 | 6080.00 |
| Federación de Rusia | RUS | 54964.00 | 55485.00 | 56527.00 | 55148.00 | 55494.00 | 57473.00 |
| Serbia | SRB | 42328.00 | 43459.00 | 43808.00 | 43882.00 | 44018.00 | 44167.00 |
| Singapur | SGP | 46411.00 | 46445.00 | 46547.00 | 45801.00 | 44232.00 | 44855.00 |
| República Eslovaca | SVK | 9511.00 | 9442.00 | 9244.00 | 9209.00 | 8791.00 | 8640.00 |
| Eslovenia | SVN | 43571.00 | 44266.00 | 45143.00 | 44772.00 | 43132.00 | 44804.00 |
| Sudáfrica | ZAF | 16068.00 | 16737.00 | 17244.00 | 17627.00 | 17560.00 | 17679.00 |
| España | ESP | 19149.00 | 20800.00 | 21965.00 | 22082.00 | 20817.00 | 21676.00 |
| Sri Lanka | LKA | 2216.00 | 2289.00 | 2384.00 | 2352.00 | 2346.00 | 2380.00 |
| Suecia | SWE | 38324.00 | 39787.00 | 41307.00 | 41999.00 | 42203.00 | 44278.00 |
| Suiza | CHE | 16002.00 | 16397.00 | 16697.00 | 17145.00 | 15894.00 | 15663.00 |
| Tailandia | THA | 6096.00 | 6228.00 | 6581.00 | 7032.00 | 7156.00 | 6760.00 |
| Ucrania | UKR | 26721.00 | 28578.00 | 30341.00 | 28788.00 | 27332.00 | 28665.00 |
| Reino Unido | GBR | 23331.00 | 24710.00 | 26406.00 | 27358.00 | 25034.00 | 26870.00 |
| Estados Unidos | USA | 55864.00 | 56613.00 | 57783.00 | 55961.00 | 53399.00 | 54449.00 |
| Uruguay | URY | 40442.00 | 41443.00 | 42485.00 | 43309.00 | 42954.00 | 44001.00 |
| Uzbekistán | UZB | 38301.00 | 38773.00 | 38969.00 | 39948.00 | 39762.00 | 39762.00 |
| Viet Nam | VNM | 24013.00 | 24606.00 | 23086.00 | 22946.00 | 23111.00 | 23681.00 |

## Bibliografía